\documentclass[11pt]{article}
\usepackage{moriond,epsfig}

\def\NIMA{{\em Nucl. Instrum. Methods} A}

\def\PRL{\em Phys. Rev. Lett.}
\def\PRD{{\em Phys. Rev.} D}

\def\be{\begin{equation}}
\def\ee{\end{equation}}
\def\bea{\begin{eqnarray}}
\def\eea{\end{eqnarray}}

\newcommand{\bsubs} {\ensuremath{B_s^0}}
\newcommand{\bsubd} {\ensuremath{B_d^0}}
\newcommand{\dzero} {D\O}
\newcommand{\unitps}  {\mathrm{ps}^{-1}}

\newcommand{\ket}[1]{|#1\rangle}

\begin{document}
\vspace*{4cm}

\title{MEASUREMENTS OF \bsubs-OSCILLATIONS AT THE TEVATRON}

\author{G.~A.~WEBER}

\address{
Johannes Gutenberg-Universit\"at Mainz\\
On behalf of the CDF and \dzero\ Collaborations. %%\textbackslash\textbackslash
}

\maketitle\abstracts{
Measuring the oscillation frequency in the \bsubs-meson system 
was one of the main goals in b-physics for
the two experiments CDF and D\O\ at the Tevatron-Collider since the start of
RunII in the year 2001.
The D\O\ collaboration was the first experiment, which was able to give a
two sided limit for the oscillation frequency of \bsubs\
	mesons. Shortly after the CDF collaboration confirmed
	this result and was able to give a $5\sigma$ measurement.
}

\section{Introduction}
The CKM matrix describes the relation between weak and flavor eigenstates of the
quarks. One of the least known matrix elements\cite{bib:pdg} is
$|V_{td}|=(7.4\pm0.8)\cdot 10^{-3}$. It is
accessible by studying the transition of neutral $B$ mesons to their anti-particles and
vice versa. This behavior is also known as mixing. It is caused by the mass
difference $\Delta m = m_H -m_L$ of the two mass eigenstates $\ket{B_H} =
p\ket{B_0}
- q\ket{\bar{B_0}}$ and  $\ket{B_L} = p\ket{B_0} + q\ket{\bar{B_0}}$.
Since the first observation of flavor oscillations in the \bsubd-system by
the ARGUS collaboration this topic has been studied intensively and was well
measured at the $B$ factories BaBar and Belle (PDG world-average\cite{bib:pdg}: $\Delta  m_d =
  \left(  0.507\pm 0.005\right)\,\unitps$).
Theoretically the relation between $\Delta m_d$ and $|V_{td}|$
is given by\cite{bib:lqcd}:
\begin{equation}
%%  \Delta m_d = \frac{G_F^2}{6\pi^2}m_{B_d^0}m_t^2
%%  F\left(\frac{m_t^2}{m_W^2}\right) B_{B_d^0} f^2_{B_d^0} |V_{tb}^* V_{td}|^2\eta_{QCD} 
  \Delta m_d =  \mathrm{(known\hphantom{\ }factor)} \times f^2_{B_d^0} B_{B_d^0} |V_{tb}^* V_{td}|^2 
\end{equation}
Nonpertubative QCD effects are contained in $f^2_{B_d^0} B_{B_d^0}$, where $f^2_{B_d^0}$ is the \bsubd\ meson decay constant and
$B_{B_d^0}$ is the \bsubd\ meson bag parameter. As there are large uncertainties in the order of 20\% on these hadronic correction
terms the determination of $ V_{td}$ is not trivial. By
measuring the mass differences $\Delta m_d$ and $\Delta m_s$ and calculating
their ratio most of the uncertainties cancel out:
\begin{equation}
\frac{\Delta m_s}{\Delta m_d} =
\frac{m_{\bsubs}}{m_{\bsubd}}\xi^2\frac{|V_{ts}|^2}{|V_{td}|^2}
\end{equation}
where $\xi^2=(1.210^{+0.047}_{-0.035})^2$ has uncertainties\cite{bib:lqcd} in
	the order of 4\%.

\section{Detectors and Datasets}
The Tevatron collider at the Fermi National Accelerator Laboratories is currently
the only place worldwide to study mixing in the \bsubs-system. Protons and
Anti-Protons collide at $\sqrt{s}$=1.96\,TeV with a bunch spacing of
396\,ns.
\subsection{The CDF II detector}
The CDF II detector\cite{bib:cdfdetector} is described in detail elsewhere. Only
parts relevant for the analysis presented will be described here. The
tracking system resides in a 1.4\,T axial magnetic field and consists of a
silicon micro-strip detector surrounded by an open-cell wire drift chamber. The muon detectors are the central muon drift
chambers, covering the pseudo-rapidity range $|\eta| < 0.6$ and the
extension muon drift chambers, covering $0.6 < |\eta| < 1.0$, where the
pseudo-rapidity is defined as $\eta = -\ln[\tan\frac{\theta}{2}]$ and $\theta$
is the polar angle.
The results\cite{bib:cdf5sigma} presented correspond to an integrated luminosity of
	$\int{\cal L}dt = 1\mathrm{fb}^{-1}$.
\subsection{The \dzero\  detector}
The \dzero\  detector\cite{bib:detector} is described in detail elsewhere. 
For the
analysis presented here the excellent muon chamber coverage in
pseudo-rapidity
$\eta$ up to $|\eta| < 2$ is advantageous and allows to collect large samples of semileptonic B
decays. Besides that the central tracking system is one of
the most important components. It consists of a silicon micro-strip tracker, a central fiber tracker and a 2\,T solenoid magnet.
The tracking system provides charged particle tracking up to  $|\eta| < 3$.
The \dzero\ results\cite{bib:combination} are based on analyses using a dataset
of an integrated luminosity of $\int{\cal L}dt = 2.4\mathrm{fb}^{-1}$.

\section{Physics Analysis}
The \bsubs-mesons are produced during the fragmentation process of $b\bar{b}$ pairs created in 
$p\bar{p}$ collisions.  To measure the oscillation frequency
the flavor at production time ($\bsubs$ or $\bar{\bsubs}$) has to be determined.
After a short distance of flight the meson decays. Hadronic and semileptonic decays are reconstructed.
It is mandatory to also know the
flavor of the \bsubs-meson at the decay vertex to determine, whether the
meson has oscillated or not. The lifetime of the \bsubs-meson can be
obtained by measuring its decay length and momentum. 
With the help of these variables as input it is possible to do a likelihood
fit and determine $\Delta m_s$.

\subsection{Event Selection}
\begin{figure}[ht]
\begin{center}
  \includegraphics[width=0.35\textwidth]{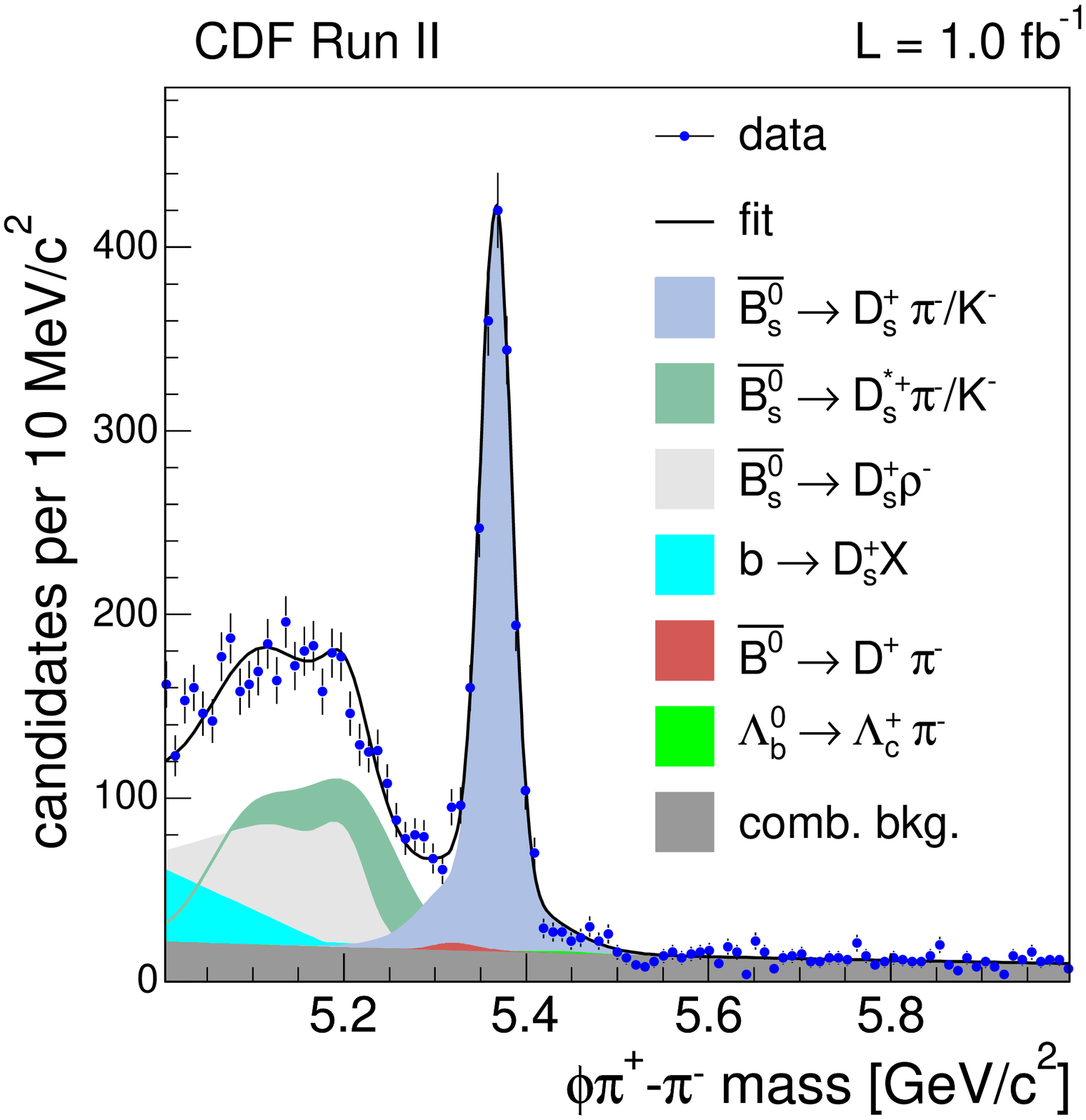}
  \includegraphics[width=0.5\textwidth]{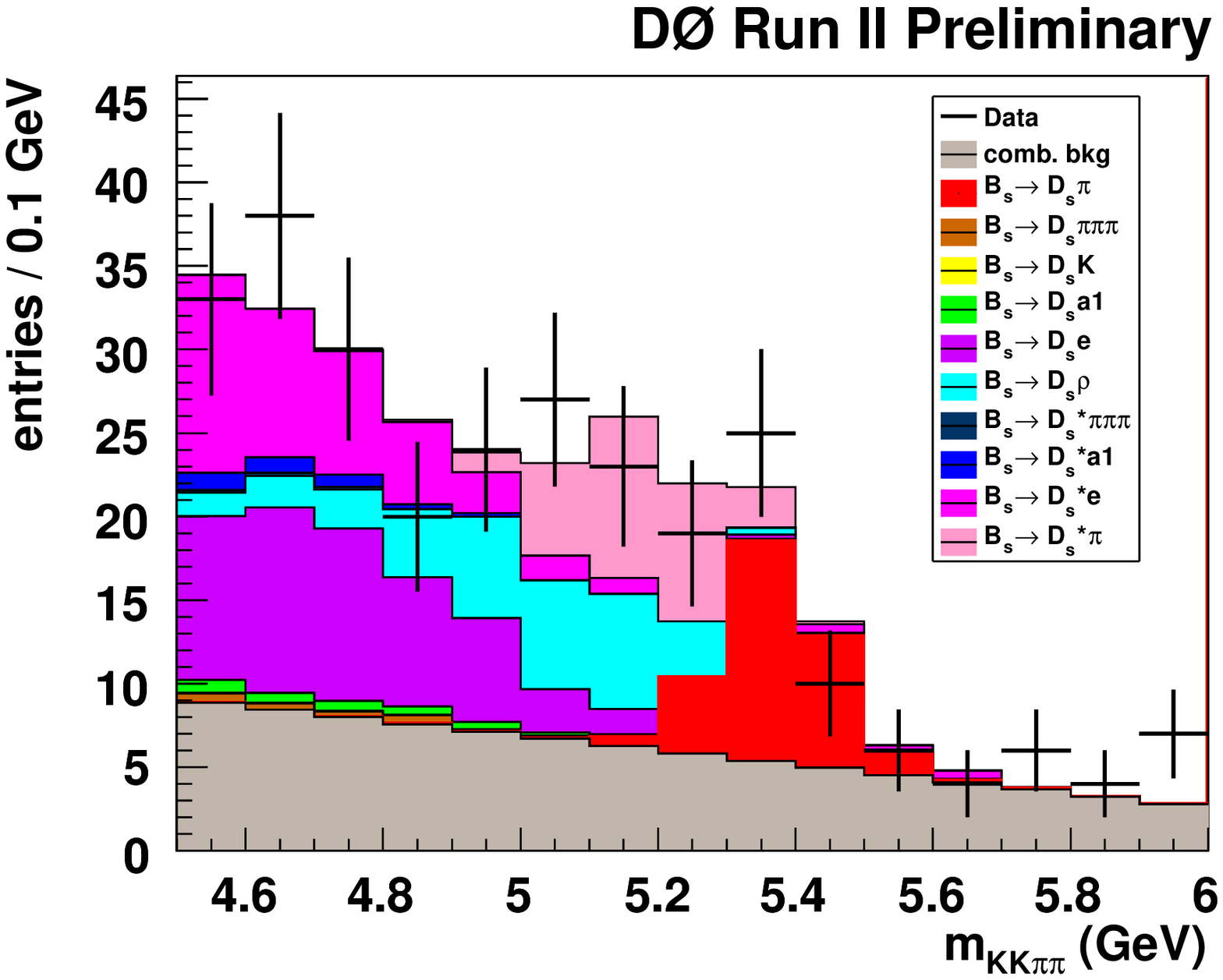}
	\caption{\label{fig:signalselection} Signal selection for the hadronic
		modes. The invariant $\bar{\bsubs}$ mass as from CDF (left) and the
			invariant \bsubs\ mass as from \dzero\ (right) is shown.} 
\end{center}
\end{figure}

The analysis starts with the reconstruction of the final state of the \bsubs\
	meson (Figure \ref{fig:signalselection}).  At \dzero\ a cut based selection is used in combination with a
	likelihood selection to obtain the final candidates. The combination of all
	semileptonic channels analyzed results in 46.5k events and for the hadronic modes
	in $249\pm 17$ events.  The
	selection done by the CDF collaboration utilizes an artificial neural
	network. This results in 61.5k semileptonic, 5.6k fully reconstructed
	hadronic and 3.1k partially reconstructed hadronic events.

\subsection{Flavor Tagging}
Flavor Tagging is an important tool to determine the initial state of the $B$ meson, i.e.
whether it is a $B$ or $\bar{B}$ and thus to clarify whether it has
oscillated or not. The tagging  power $\epsilon\mathcal{D}^2$ is described by the combination of the two
quantities efficiency and dilution.
The efficiency $\epsilon$ is the fraction of reconstructed
$B$ mesons, that are tagged, divided by the total number:
$\epsilon = \frac{N_{tag}}{N_{tot}}$.
The dilution $\mathcal{D}$ is given by
$\mathcal{D} = 2\eta-1$, 
where the purity $\eta$ is given by the fraction of correctly tagged events
divided by the total number of tagged events $\eta = \frac{N_{cor}}{N_{tag}}$.
The tagging is done by either observing the decay particles on the
opposite side (OST: opposite side tagging) or by studying the decay side
itself (SST: same side tagging). The \dzero\ collaboration uses \bsubd\ data
to determine the tagging power and obtains a value of $\epsilon{\cal D}^2 =
(2.48\pm 0.021_{stat} \hphantom{x}^{+ 0.08}_{-0.06})\%$ for OST and  $\epsilon{\cal
	D}^2 = (2.2\pm 0.1)\%$ (semileptonic modes) for SST. No SST is used for the
	hadronic modes at \dzero. The OST power of the CDF experiment is
	slightly smaller: $\epsilon{\cal D}^2 =
	(1.8\pm 0.1)\%$. But for the SST an artificial
	neural network is used in CDF, which is trained with Monte Carlo samples and tested with \bsubd\
	samples. The tagging power for this technique exceeds the \dzero\ method
	and results in $\epsilon{\cal D}^2 =
   (3.7\pm 0.9)\%$ for the hadronic modes and $\epsilon{\cal D}^2 =
   (4.8\pm 1.2)\%$ for the semileptonic modes.

\subsection{Proper Decay Length Measurement}
To determine the decay time the flight length in the transverse plane ($L_{xy}^B$) and the momentum of the
\bsubs-meson has to be measured:
\begin{equation}
c t_{B_s^0} = L_{xy}^B\cdot\frac{M_{B_s}^0}{p_T(B_s^0)}\cdot K, 
	\mathrm{\ }K \equiv \frac{p_T(B_\mathrm{meas.})}{p_T(B_s^0)},
\end{equation}
%% \begin{equation}
%% c t_{B_s^0} =L_{xy}^B\cdot\frac{M_{B_s}^0}{p_T(B_s^0)}\cdot K\mathrm{, with\
%% }K\equiv\frac{p_T(B_\mathrm{meas.})}{p_T(B_s^0)},
%% \end{equation}
where $p_T(B_\mathrm{meas.})$ is the measured transverse momentum determined
through all detected particles in contrast to the true transverse momentum of
the \bsubs-meson $p_T(\bsubs)$. For fully reconstructed events the measured momentum is equal to the true momentum and so $K\equiv
1$. The correction term for not fully reconstructed decays is determined through
Monte Carlo studies.

\section{Results and Conclusions}
\begin{figure}[ht]
  \includegraphics[width=0.5\textwidth]{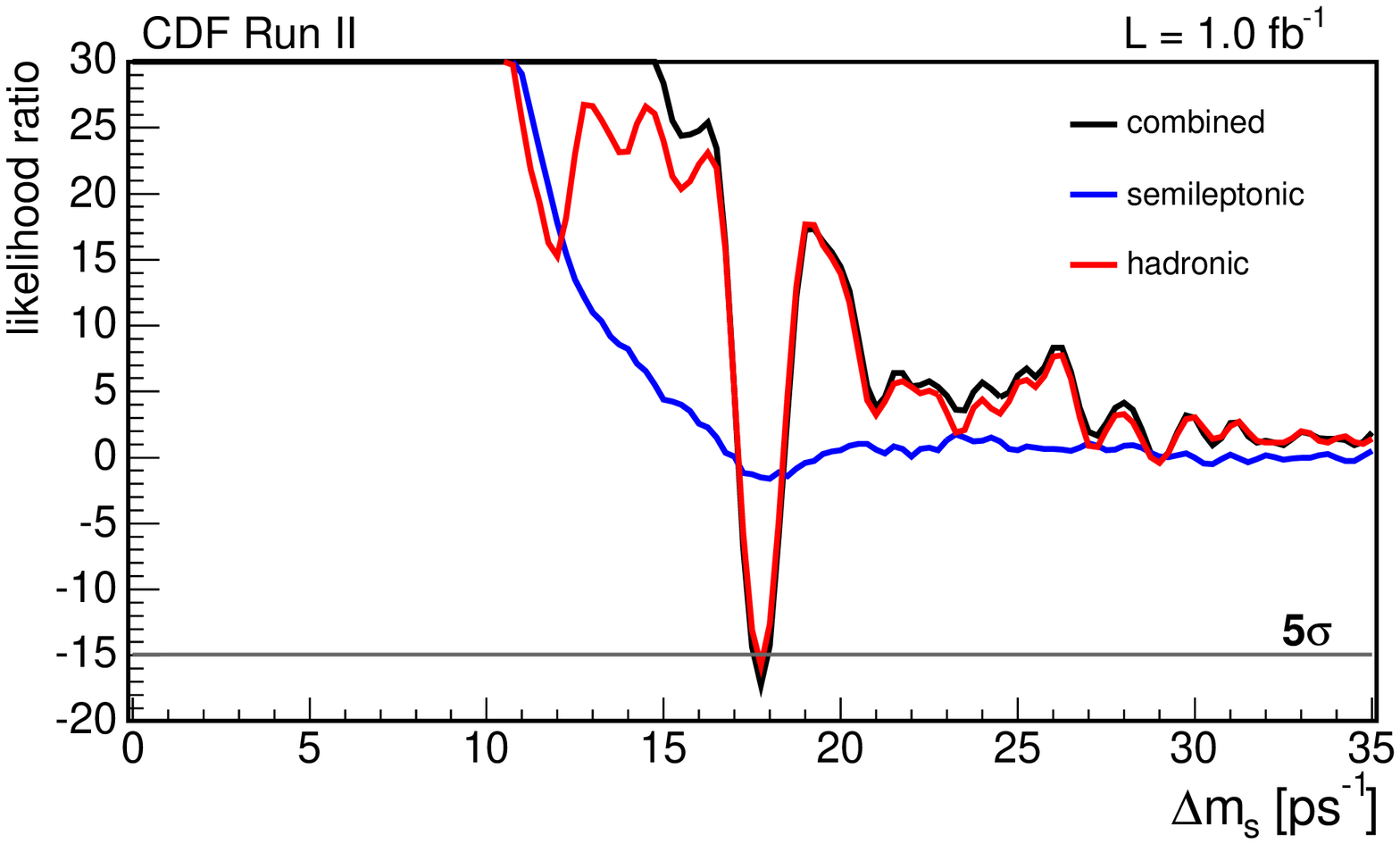}
  \includegraphics[width=0.5\textwidth]{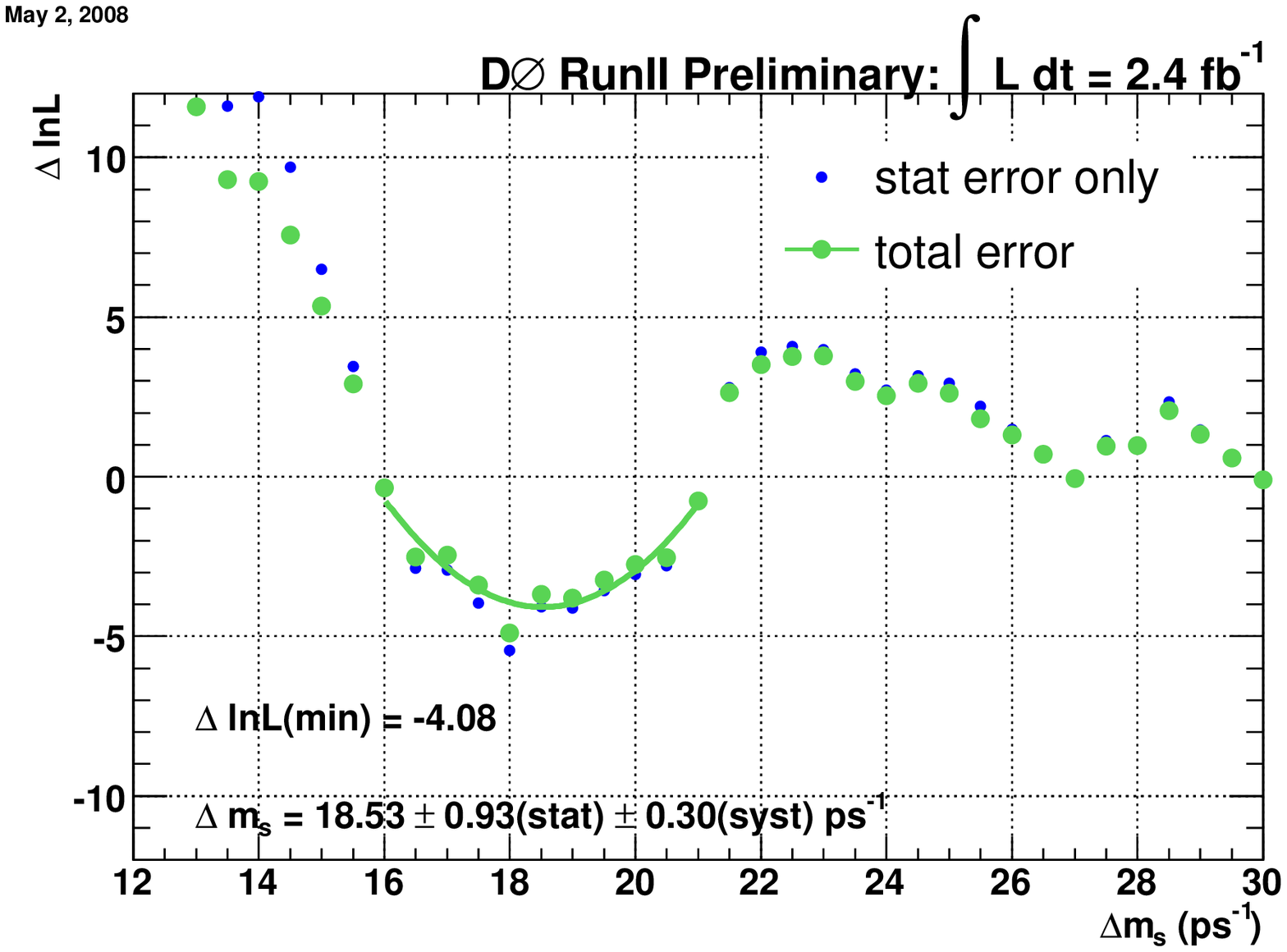}
	\caption{\label{fig:likelihood} Log-likelihood fit for CDF (left) and
		\dzero\ (right) data.} 
\end{figure}

Using an integrated luminosity of $\int{\cal L}dt = 1\mathrm{fb}^{-1}$ the
CDF collaboration measured the \bsubs-oscillation frequency as
\begin{equation}
\Delta m_s = (17.77\pm 0.10_{stat}\pm 0.07_{sys})\unitps,
\end{equation}
while the \dzero\ experiment used a larger dataset of $\int{\cal L}dt =
2.4\mathrm{fb}^{-1}$ and obtained a value of
\begin{equation}
\Delta m_s = (18.53\pm 0.93_{stat}\pm 0.30_{sys})\unitps.
\end{equation}
Figure \ref{fig:likelihood} shows the distribution of $\Delta \mathrm{log}
{\cal L}$ as a function of $\Delta m_s$ for both experiments, where the lowest data point equals
to the most probable value for $\Delta m_s$.

\noindent Therefore the ratio $\frac{|V_{td}|}{|V_{ts}|}$ can be calculated as
\begin{eqnarray}
\mathrm{CDF:\hphantom{\ }} \frac{|V_{td}|}{|V_{ts}|}&=&0.2060\pm 0.0007(exp)^{+0.0081}_{-0.0060}(theor) \\
\mathrm{\dzero:\hphantom{\ }} \frac{|V_{td}|}{|V_{ts}|}&=&0.2018\pm
0.005(exp)^{+0.0079}_{-0.0059}(theor).
\end{eqnarray}
The obtained values are now dominated by the theoretical uncertainties and in
perfect agreement with the standard model predictions\cite{bib:lqcd} of $\frac{|V_{ts}|}{|V_{ts}|}_{Lat05} = 0.1953$. The present
measurements exclude contributions of new physics amplitudes exceeding in size the standard model contributions.

\section*{Acknowledgments}
I would like to thank the staff at Fermilab and collaborating institutions,
  and acknowledge support from the BMBF (under contract 05HA6UMA). Special thanks
  must be extended to the organizers of this pleasant conference.

\end{document}